\documentclass[]{elsarticle}
\pdfoutput=1
\usepackage{todonotes}
\usepackage{url}
\usepackage{listings}
\usepackage{subcaption}
\usepackage[htt]{hyphenat}

\begin{document}

\title{ds-array: A Distributed Data Structure for Large Scale Machine Learning}

\author[bsc]{Javier Álvarez Cid-Fuentes}
\author[pol]{Pol Álvarez}
\author[bsc]{Salvi Solà}
\author[fuji]{Kuninori Ishii}
\author[fuji]{Rafael K. Morizawa}
\author[bsc]{Rosa M. Badia}

\address[bsc]{Barcelona Supercomputing Center (BSC)}
\address[pol]{Pensionera Förmedling i Sverige AB}
\address[fuji]{Fujitsu Ltd.}

\begin{abstract}
Machine learning has proved to be a useful tool for extracting knowledge from scientific data in numerous research fields, including astrophysics, genomics, and molecular dynamics. Often, data sets from these research areas need to be processed in distributed platforms due to their magnitude. This can be done using one of the various distributed machine learning libraries available. One of these libraries is dislib, a distributed machine learning library for Python especially designed to process large scale data sets on HPC clusters, which makes dislib an ideal candidate for analyzing scientific data. However, dislib's main distributed data structure, called Dataset, has some limitations, including poor performance in certain operations and low flexibility and usability. In this paper, we propose a novel distributed data structure for dislib, called ds-array, that addresses dislib's main limitations in data management. Ds-arrays simplify distributed data management in dislib by exposing a NumPy-like API, provide more flexibility, and reduce the computational complexity of some operations. This results in performance improvements of up to two orders of magnitude over Datasets, while also greatly improving scalability and usability.
\end{abstract}

\maketitle

\section{Introduction}

Python is one of the most popular programming languages among developers~\cite{cass19} because it is easy to learn, and provides high productivity and a rich ecosystem of libraries and tools. Python is also an effective programming language for scientific applications~\cite{beazley97,hinsen97}, and has become very popular among scientists. Especially after the appearance of numerical libraries, like NumPy~\cite{walt11} and pandas~\cite{mckinney10}, and machine learning libraries like scikit-learn~\cite{pedregosa11}. 

Scikit-learn made machine learning accessible to non-experts, and has allowed researchers from different fields to apply machine learning to their data sets. Examples of research areas where machine learning has proved useful are astrophysics~\cite{castro20}, molecular biology~\cite{nauman19}, genomics~\cite{libbrecht15}, and medicine~\cite{qian19}. However, as scientific data sets grow in size, it appears a need for distributed machine learning libraries that can run in traditional computational science platforms like HPC clusters. Towards this, some machine learning libraries, like MLlib~\cite{meng16}, Dask-ML~\cite{daskml}, dislib~\cite{alvarez19}, and TensorFlow~\cite{abadi16} have addressed scikit-learn's limitations by being able to run in multiple computers. Among these libraries, dislib is one of the better suited for HPC clusters, as it provides better performance and scalability than other similar libraries when processing large data sets in these environments~\cite{alvarez19}. Dislib is built on top of PyCOMPSs programming model~\cite{lordan14,tejedor17}, and exposes two main APIs: a data handling interface to manage large scale data as if it was stored locally, and an estimator-based interface that provides the various machine learning models included in the library in an easy-to-use manner. 

Although dislib is an efficient and scalable distributed library, its data handling API has some limitations. This API abstracts developers from the management of the data in distributed environments by means of the \emph{Dataset} structure. Datasets are Python objects that can be operated in a sequential manner, but that are internally parallelized. More precisely, Datasets are collections of samples and labels that are divided in partitions called \emph{Subsets}. Each Subset can be stored in a different computer, and contains a block of samples and a block of labels, where samples are vectors and labels are numbers. This partitioning strategy is efficient in scenarios where samples are accessed in subsets. However, it is extremely inefficient when accessing the different components of each sample vector. This has a negative impact on the performance of various operations and algorithms included in dislib. In addition to this, Datasets provide a limited API, cannot store samples and labels independently, and provide limited performance for filtering operations (i.e., fetching a group of samples from each Subset).

In this paper, we propose a novel data structure for dislib to address the limitations of Datasets that we call \emph{distributed array} or ds-array. Ds-arrays are 2-dimensional arrays (i.e., matrices) divided in blocks that can be stored in different computers. Internally, ds-arrays are stored as collections of NumPy arrays or SciPy~\cite{virtanen19} sparse matrices depending on the characteristics of the data. In an effort to make ds-arrays easy to use, we have implemented a NumPy-like interface that allows different types of indexing, and other mathematical operations like the mean and the transpose, which are automatically parallelized using PyCOMPSs.

Contrary to Datasets, ds-arrays do not require data to be divided into samples and labels. Instead, in a similar way to NumPy arrays, ds-arrays can represent any kind of numerical data. This makes ds-arrays a more flexible data structure than Datasets, and simplifies dislib's estimator-based API. Moreover, unlike Datasets, ds-arrays can be divided in blocks of an arbitrary size. This improves the performance of data access patterns that are inefficient with Datasets, such as accessing the different components of each sample (i.e., traversing a ds-array along the second dimension). Ds-arrays also simplify dislib's data handling interface due to their NumPy-like API, and achieve significant improvements in the performance of certain operations due to a more efficient implementation that takes advantage of some of PyCOMPSs' latest features.

The rest of this paper is organized as follows: Section~\ref{sec:related} overviews similar works in the area. Section~\ref{sec:background} contextualizes the contribution of this paper by providing the necessary background. Section~\ref{sec:array} presents this contribution. Section~\ref{sec:evaluation} presents a performance comparison between Datasets and ds-arrays, and Section~\ref{sec:conclusions} concludes the paper.

\section{Related work}
\label{sec:related}

In this section, we list previous works that are similar to ds-arrays in the sense that they provide distributed arrays for Python through a NumPy-like API.

Global Arrays in NumPy (GAiN)~\cite{daily11} provides a NumPy-like API on top of the Global Arrays (GA) toolkit~\cite{nieplocha06}. GAiN re-implement the whole NumPy module, and thus serve as a drop-in replacement for NumPy. GAiN is similar to ds-arrays, but is parallelized with MPI~\cite{gropp96} and uses the GA toolkit as backend for storing data, while ds-arrays are parallelized with PyCOMPSs and use NumPy arrays and SciPy sparse matrices as backends.

Spartan~\cite{huang15} is another drop-in replacement for NumPy that puts a strong emphasis on smart tiling, that is, deciding the most appropriate array partitioning based on the operations that need to be computed. Spartan achieves this by deferring computations. Given a series of array operations, Spartan first creates a computation graph of high-level operators, then runs an optimizer to find the best tiling for the given graph, and finally runs the computations on the available resources. The main priority of the optimizer is to minimize data transfers. The main difference between ds-arrays and Spartan is that ds-arrays are built on top of an existing distributed programming model, while Spartan is an ad-hoc distributed array solution.

Dask~\cite{dask} is a framework for parallel and distributed computing in Python that provides various parallel data structures. One of these structures is the Dask array, which exposes an API similar to NumPy. Dask transforms a sequence of array operations into a graph task, and then schedules the tasks in the available resources following a master-worker approach. Internally, Dask arrays are collections of NumPy arrays that can be stored in different computers. Dask runtime is similar to PyCOMPSs in that both build task graphs and follow a master-worker approach. The main difference between ds-arrays and Dask arrays is that Dask delays the computation of the task graph until an actual result is needed, while ds-arrays submit tasks for execution as soon as operations are carried out. Apart from this, ds-arrays support SciPy sparse matrices as array partitions.

The Helmholtz Analytics Toolkit (HeAT)~\cite{krajsek18} provides distributed arrays (or tensors) for Python based on PyTorch~\cite{pytorch} and MPI. The motivation behind HeAT tensors is the difficulty in managing PyTorch tensors in distributed environments, such as HPC clusters. HeAT tensors are n-dimensional arrays divided in one or more PyTorch tensors that can be stored and processed using multiple computers. HeAT tensors expose a sequential NumPy-like API, and parallelism and data distribution is automatically handled by HeAT using MPI. Even though HeAT tensors are n-dimensional, they can only be distributed along the first or the second dimension. HeAT supports both CPU and GPU computations, and also provides some machine learning models, such as K-means, directly implemented on top of HeAT tensors. The main differences between the work presented in this paper and HeAT are that ds-arrays are parallelized with PyCOMPSs while HeAT tensors are parallelized with MPI; and that HeAT tensors are built on top of PyTorch while ds-arrays are built on top of NumPy arrays and SciPy sparse matrices.

Legate~\cite{bauer19} is one the most recent implementations of distributed arrays for Python. Legate is built on top of the Legion task-based programming model~\cite{bauer12}, and provides a drop-in replacement for NumPy. Since Legion consists of a data-driven runtime system that organizes data into distributed logical regions, Legate can be seen as a NumPy abstraction on top of Legion logical regions. Thus, Legate maps arrays to logical regions, and translates array operations to Legion tasks that are executed asynchronously. These tasks are programmed in CUDA or C++, and are executed in the available nodes leveraging system libraries like OpenBLAS~\cite{wang13} and cuBLAS~\cite{cublas}, and parallel programming models like OpenMP~\cite{dagum98}. One important aspect of Legate is that it provides support for clusters of GPUs and CPUs, which can speed up computations greatly. The main difference between Legate and ds-arrays are that ds-arrays are backed up directly with NumPy arrays and SciPy sparse matrices instead of Legion logical regions.

\section{Background}
\label{sec:background}

This section provides a brief overview on PyCOMPSs and dislib to contextualize the contribution of this paper. The reader can obtain additional details on these two systems in Lordan et. al~\cite{lordan14}, Tejedor et. al~\cite{tejedor17}, and Álvarez Cid-Fuentes et. al~\cite{alvarez19}.

\subsection{PyCOMPSs framework}

PyCOMPSs is a programming model and runtime that simplifies the development and execution of parallel applications in distributed environments. PyCOMPSs has proven to be an efficient framework for data analytics~\cite{alvarez19b}. PyCOMPSs' programming model is mainly based on the \emph{task} annotation, which can be used to define units of computation in the application. PyCOMPSs' runtime analyzes the source code of the annotated application, and distributes these units of computation among the available computational nodes. PyCOMPSs' runtime does this by taking into account data dependencies between tasks to exploit the potential parallelism of the application.

\subsubsection{Programming model}

A PyCOMPSs application consists of two main parts: main code and task definitions. The main code of the application is regular Python code that calls different functions defined as tasks. To define a function as a task, PyCOMPSs provides the \texttt{@task} Python decorator. Figure~\ref{fig:annotation} shows an example of main code (line 1) and task definition (line 7) of a simple PyCOMPSs program.

\begin{figure}[h]
\centering
\begin{lstlisting}
def main:
    s = 0

    for i in range(10):
        s += multiply(i, i)
        
@task(num1=IN, num2=IN, returns=int)
def multiply(num1, num2):
    return num1 * num2
\end{lstlisting}
\caption{Example PyCOMPSs application.}
\label{fig:annotation}
\end{figure}

The \texttt{@task} decorator can take various arguments that PyCOMPSs' runtime uses during the execution of the application. The most important of these arguments are parameter type, parameter direction, and return type. Parameter type indicates the type of a task argument (e.g., integer), parameter direction indicates whether the argument of a task is read or written within the task, and return type indicates the type of the returned value.  

\subsubsection{Runtime}

PyCOMPSs' runtime has a master-worker architecture. The master process executes the main code of the application, while workers execute the different tasks. Typically, master and workers run on different computational nodes. The master process inserts a new task into a dependency graph every time that the main code of the application calls a function that has been annotated as a task. At the same time, the master process continuously monitors this dependency graph, and sends dependency-free tasks to available workers for execution. A task becomes dependency-free when all of its inputs are available. PyCOMPSs' runtime infers data dependencies between tasks by looking at the parameter directions in the task decorators.

Task scheduling is asynchronous. This means that the master process does not wait for the results of a task unless there is an explicit synchronization point in the main code of the application. In Figure~\ref{fig:annotation} (line 5), \texttt{s} does not contain the actual value returned by \texttt{multiply}. Instead, \texttt{s} is a placeholder or future object. To retrieve the actual contents of a future object, developers can use the \texttt{compss\_wait\_on(object)} call.

\subsection{dislib}
\label{sec:dislib}

Dislib~\cite{alvarez19} is a distributed machine learning library built on top of PyCOMPSs programming model. In essence, dislib is a collection of PyCOMPSs applications exposed through two main APIs: an estimator-based interface and a data handling interface. The estimator-based interface is inspired by scikit-learn and provides easy access to the different machine learning algorithms included in dislib. An estimator is anything that learns from data. For example, a K-means estimator \emph{learns} the optimal cluster centers given a collection of samples. Dislib's data handling interface provides functions to manage distributed data in a simple manner. This includes loading data from disk in parallel, and reading and writing this data in a way that abstracts developers from manually handling data partitions and transfers.

All dislib estimators and functions are internally parallelized with PyCOMPSs. This means that calls to dislib operators generate multiple tasks that are executed in the available workers. In the following sections, we describe dislib's interfaces in more detail.

\subsubsection{Data management}

\begin{figure}[h]
	\centering
		\includegraphics[width=0.3\linewidth]{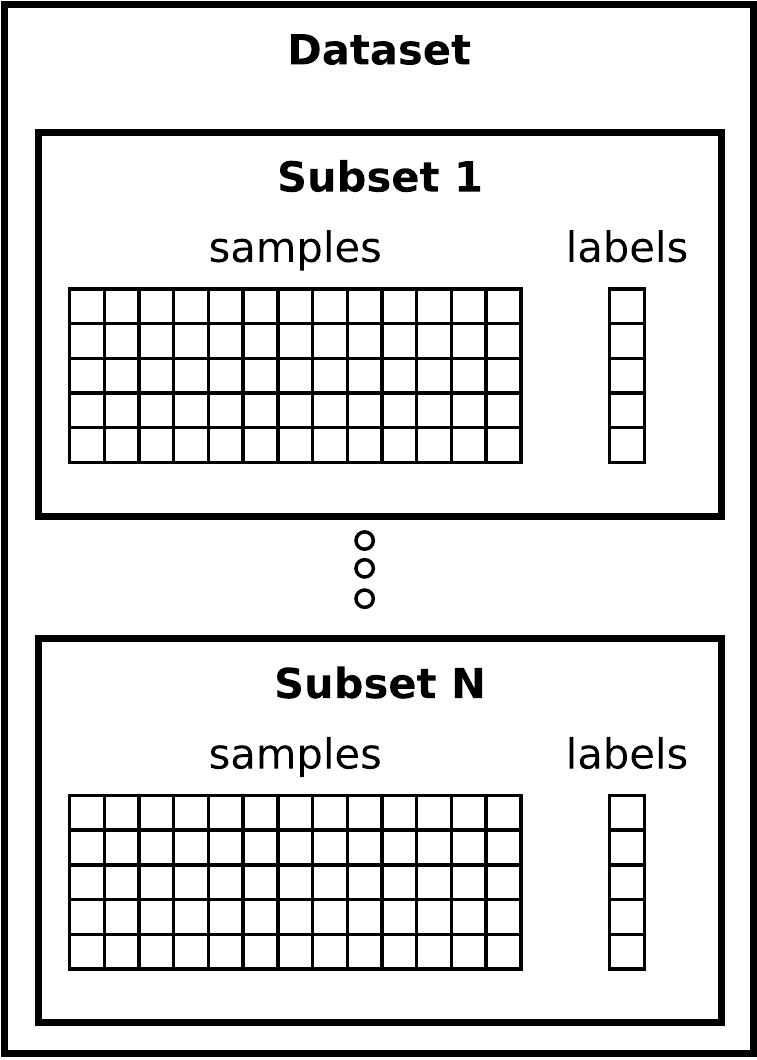}	
	\caption{Internal structure of a Dataset.}
	\label{fig:dataset}
\end{figure}

Dislib's data handling interface is based on the Dataset and Subset classes. A Dataset is a collection of samples and labels, where each sample is an n-dimensional vector, and each label is a number that represents the category of a sample. Datasets are divided in Subsets that can be stored remotely. Internally, Subsets store a set of samples and labels as two NumPy arrays of size $N \times M$ and $N \times 1$ respectively, where $N$ is the number of samples and $M$ is their dimensionality (i.e., the number of features). Samples can also be stored as SciPy CSR matrices in the case of sparse data. Figure~\ref{fig:dataset} shows the internal structure of a Dataset.

Dislib also provides routines to load data from one or multiple files into a Dataset. These routines create one PyCOMPSs task per Subset, and load the contents of the file/s in parallel. In the case of loading data from a single file, users can decide the size of the Subsets. Loading data in parallel using PyCOMPSs tasks means that Subsets are not stored in local memory, but remotely and represented with future objects. Nevertheless, users can interact with Dataset objects as if they were stored locally. In order to retrieve the actual contents of a Dataset, users can access class attributes \texttt{samples} and \texttt{labels}. This synchronizes the remote data and returns all the samples or labels in the Dataset as NumPy arrays (or a SciPy CSR matrix if the data is sparse). Apart from this, Datasets provide some useful methods, such as \texttt{transpose}, \texttt{append}, \texttt{max\_features}, \texttt{min\_features}, and \texttt{subset\_size}. The \texttt{transpose} returns a new Dataset with transposed samples; \texttt{append} adds a new Subset to the Dataset; \texttt{max\_features} and \texttt{min\_features} return the maximum and minimum features of the Dataset as NumPy arrays; and \texttt{subset\_size} returns the size of a particular Subset. 

\subsubsection{Estimators}

Dislib exposes the different machine learning algorithms in the library through an estimator-based interface inspired by scikit-learn. As said before, an estimator is anything that learns from data. The typical workflow when working with dislib consists of the following steps: 1) load data into a Dataset; 2) instantiate an estimator with parameters; 3) fit the estimator with the Dataset; 4) make predictions on new data or extract information from the estimator. 

The two main methods used in this workflow are \texttt{fit(dataset)} and \texttt{predict(dataset)}. The \emph{fit} method computes estimator statistics using the samples and labels in the input Dataset. For example, the K-means estimator computes the optimal cluster centers, while the support vector machine classifier estimator computes a decision function. The \emph{predict} method typically computes the labels of unlabeled samples given a fitted estimator. Computed labels are stored in the input Dataset, and can be accessed with the \texttt{labels} attribute of the Dataset class. Other useful information computed in the \emph{fit} method is stored as future objects in the estimator class. Examples of this are the total number of clusters found or the support vectors themselves. Both \emph{fit} and \emph{predict} methods are internally parallelized with PyCOMPSs. In most cases, this means that dislib processes each Subset in parallel, and eventually performs some kind of reduction to get the final results, or to decide whether the algorithm has converged.

\section{Distributed arrays}
\label{sec:array}

In this section, we analyze the current limitations of data management in dislib, and propose the distributed array data structure to overcome these limitations.

\subsection{Current limitations}
\label{sec:limitations}

The main limitations of dislib's data management are in performance and usability. The Dataset structure is useful for data that is divided into samples and labels, which is a common representation in machine learning. However, in many cases, this representation is not the most convenient, and a more general distributed data representation is required. For example, the alternating least squares (ALS) algorithm~\cite{hu08} receives a matrix as input instead of a set of samples and labels. The current implementation of the \texttt{fit} method of ALS in dislib takes a Dataset as input parameter, and treats the samples as the input matrix while ignoring the labels. This is an abuse of the Dataset structure that would be avoided with a more general distributed data structure. 

In addition to this, machine learning algorithms that treat the input samples as a matrix, such as ALS, often perform matrix operations as part of the fitting process. This includes matrix multiplication, matrix decomposition, or the transpose. These operations are not easy to define in the Dataset abstraction as it is not clear what the resulting labels should be. Moreover, some machine learning algorithms might need to perform certain operations on the labels instead of on the samples. The Dataset abstraction can only provide this type of functionality by duplicating methods in its API, such as by providing \texttt{transpose\_labels} and \texttt{transpose\_samples}. This makes the Dataset API more complex, and reduces its usability.

Other cases where a more general distributed data structure would result in a more usable interface are in the \texttt{predict(dataset)} and \texttt{score(dataset)} estimator methods. These methods compute and return a value for each sample in a Dataset, and these computed values need to be stored in a distributed way as they might not fit in memory. Since the only distributed data structure in dislib is the Dataset, \texttt{predict} and \texttt{score} can be implemented in two ways: returning a new Dataset containing the computed values in the labels field (with empty samples), or storing the computed values directly in the input Dataset. Currently, dislib implements the second option. However, both solutions are counter-intuitive to the developer and an abuse of the Dataset class.

\begin{figure}[h]
	\centering
		\includegraphics[width=0.6\linewidth]{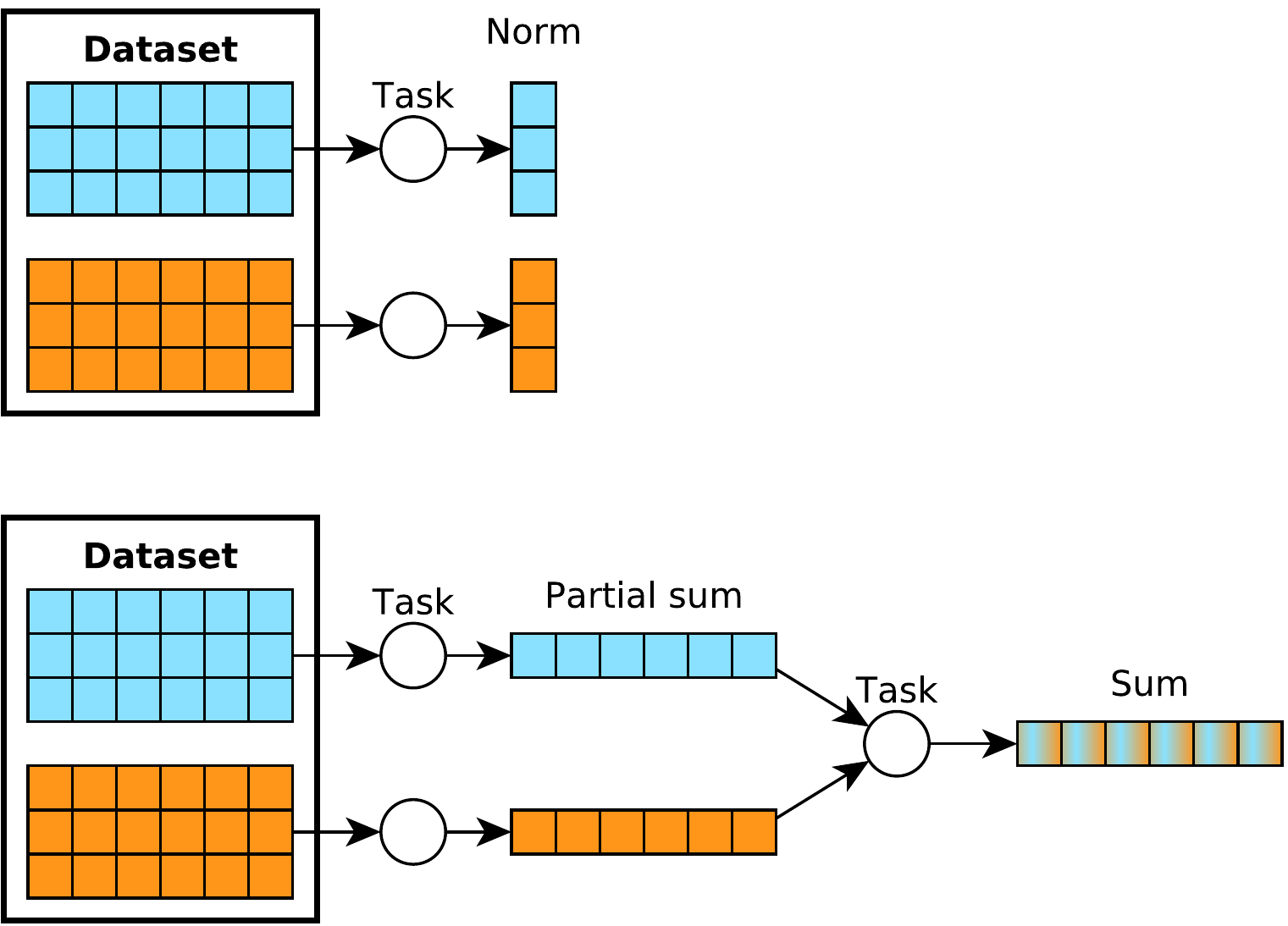}	
	\caption{Computation of the norm and the summation of samples in a Dataset.}
	\label{fig:norm}
\end{figure}

Apart from these limitations in usability, the design of the Dataset class also limits its performance. On the one hand, Datasets provide parallelization across the vertical axis only. This means that operations that need to access elements in various Subsets are less efficient than operations that access samples in one Subset. For example, computing the norm of each sample can be done in parallel by creating one task per Subset. However, computing the summation of all samples requires one task per Subset and a reduction to merge the partial sums. Figure~\ref{fig:norm} illustrates the difference in the computation of the norm and the summation of samples in a Dataset.

On the other hand, the internal structure of Datasets reduces the performance of certain operations. For example, transposing the samples in a Dataset requires dividing each Subset into $N$ parts, and then merging these parts into the new Subsets of the transposed Dataset. This process requires $N^2 + N$ tasks in total, and would be much more efficient if Datasets were divided in square blocks instead of subsets of samples.

Finally, Dataset operations that redistribute samples among Subsets, such as \emph{shuffle}, currently require around $N^2$ tasks, where $N$ is the number of Subsets. This could be improved by using some of PyCOMPSs latest features that were not available when Datasets were designed.

\subsection{Proposed design}

\begin{figure}[h]
	\centering
		\includegraphics[width=0.2\linewidth]{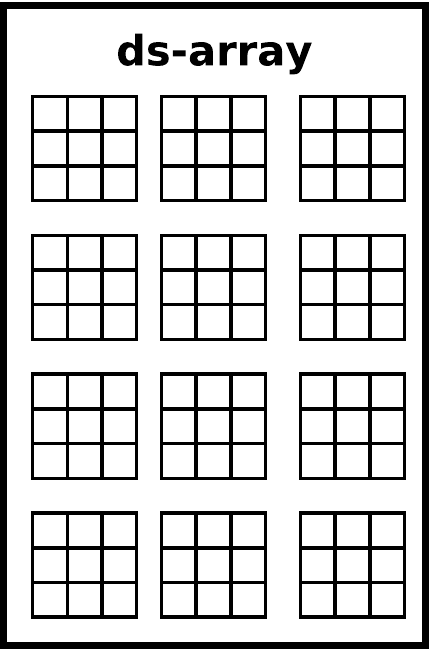}	
	\caption{Internal representation of a distributed array (each 3x3 square represents a block).}
	\label{fig:array}
\end{figure}

We address the current limitations of dislib data management with a new distributed data structure called distributed array or ds-array. A ds-array is a 2-dimensional data structure that is divided in blocks. Similar to Subsets in a Dataset, blocks in a ds-array are not stored in local memory, but distributed in the memory of all the available nodes. Blocks are NumPy arrays or SciPy CSR matrices depending on the type of data used to initialize the ds-array. Figure~\ref{fig:array} shows the internal structure of a ds-array as opposed to the structure of a Dataset shown in Figure~\ref{fig:dataset}.

The design of ds-arrays does not make any assumptions on the characteristics of the data stored in them. This means that ds-arrays can be used both for samples, labels, and any other data that needs to be stored in a distributed way. For example, a set of $N$ samples of $M$ features can be represented with a ds-array of total size $N \times M$ divided in blocks of size $P \times Q$ where $P \leq N$ and $Q \leq M$.

\subsubsection{Internal implementation}
\label{sec:internal}

The ds-array data structure mainly consists of a list of lists of blocks. The block at position $(i, j)$ on this list of lists represents the block at row $i$ and column $j$ on the ds-array. Since blocks are stored remotely, developers cannot access data directly. Instead, blocks are represented with future objects, and their actual is only accessible in tasks, or by explicitly retrieving the data.

In order to internally parallelize ds-arrays, we make extensive use of two data types that have been recently added to PyCOMPSs: \texttt{COLLECTION\_IN} and \texttt{COLLECTION\_OUT}. These data types can be used to define lists as input or output parameters of tasks. PyCOMPSs then treats each element of these lists as an independent object in terms of computing data dependencies between tasks. This allows the creation of tasks with a variable number of inputs and outputs, which is useful for tasks that take or generate multiple ds-array blocks.

\subsubsection{Array creation}

Like Datasets, ds-arrays can be created by reading data from a file, or by using an array creation routine, such as \texttt{random\_array}, which creates a ds-array containing random data. Array creation routines take the block size as input parameter to allow developers to manage how the array is distributed. All blocks are created with the same size except from the rightmost blocks and the blocks at the bottom, which can be smaller if the total size of the array is not divisible by the defined block size. Functions that create a ds-array from a file typically spawn one task per row of blocks because files are parsed line by line. Other functions, such as \texttt{random\_array}, might spawn one task per block. In this way, ds-arrays are created in parallel and directly stored among the available resources.

\subsubsection{Array interface}

The interface of ds-arrays is based on NumPy and includes indexing and other useful operations. This interface allows developers to interact with the ds-array as if it was stored in local memory.

Ds-arrays provide various forms of indexing that are expressed with square brackets (i.e, \texttt{A[...]}) and that are used to retrieve a subset of elements in the array. For example, \texttt{A[10:100]} retrieves rows from 10 to 100 from ds-array \texttt{A}. Ds-arrays also provide element-wise algebraic operators, such as the power operator (expressed as \texttt{A**2}), and matrix operations like the transpose or the multiplication. Apart from these, ds-arrays also provide a \texttt{collect} method that retrieves the actual contents of the array from the remote nodes. This method synchronizes and merges the different blocks of the ds-array and returns an equivalent NumPy array.

All the operations on ds-arrays are performed in parallel and return a new transformed ds-array. This allows the concatenation of multiple operations in a similar way to NumPy, and enables writing mathematical expressions that are automatically parallelized with PyCOMPSs. For example, the following Python code: 

\texttt{(w.transpose().norm(axis=1) ** 2).sqrt()} 

\noindent where \texttt{w} is a ds-array, defines the mathematical expression $\sqrt{\parallel{w^T}\parallel_{2}^{2}}$ and runs automatically in parallel.

\subsection{Addressing current limitations}

Distributed arrays address dislib limitations (see Section~\ref{sec:limitations}) in various ways. The ds-array interface is based on NumPy, which is a widely used Python library. This makes ds-arrays easy to use to people already familiar with NumPy, and facilitates the parallelization of already existing NumPy applications. 

In addition to this, ds-arrays provide a general data representation that is not limited to samples and labels. This simplifies dislib's estimator interface, and makes it more usable. The \texttt{fit(dataset)} method becomes \texttt{fit(x, y)}, where \texttt{x} and \texttt{y} are ds-arrays representing samples and labels respectively. This definition is more consistent with scikit-learn, and thus easier to understand by many Python developers. Moreover, other estimator methods, such as \texttt{score(dataset)} and \texttt{predict(dataset)}, can be defined as \texttt{score(x, y)} and \texttt{predict(x, y)}, and implemented such that they return a new ds-array that represents the computed data. This is much more intuitive than returning a Dataset with empty samples or filling up the labels of the input Dataset.

\begin{figure}[h]
	\centering
		\includegraphics[width=0.2\linewidth]{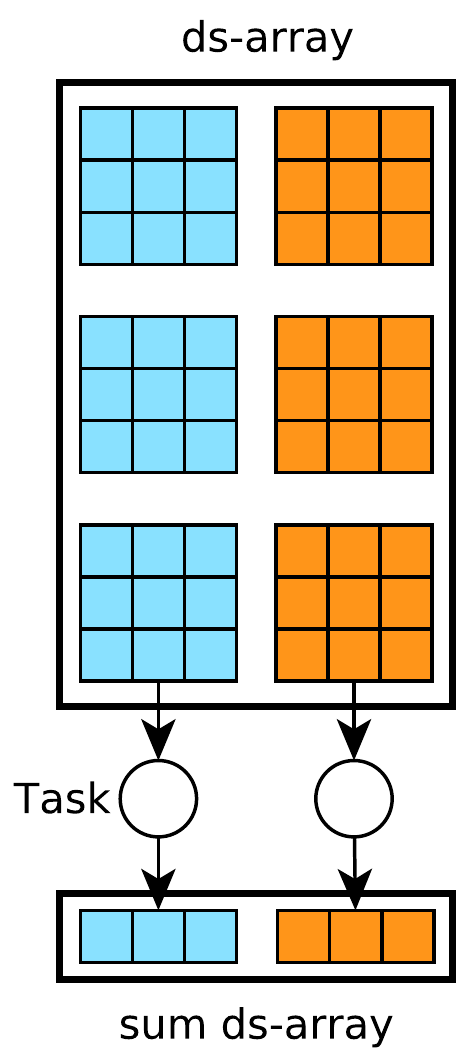}	
	\caption{Implementation of the summation of rows in a ds-array.}
	\label{fig:sum}
\end{figure}

In terms of performance, dividing ds-arrays in blocks of variable size has several advantages. On the one hand, ds-arrays provide parallelization across both axes. This improves the performance of some operations, such as the summation of rows. Figure~\ref{fig:sum} shows how this operation is implemented in ds-arrays. Each task takes three blocks (i.e., one column of blocks) to generate a sum ds-array. As it can be seen, ds-arrays allow for a more efficient implementation of this kind of operations than Datasets (see Figure~\ref{fig:norm}) due to the fact that the ds-array is divided both vertically and horizontally. An equivalent implementation with Datasets would require to load all Subsets into memory, which might not be feasible. Finally, another advantage of ds-arrays is that the resulting sum can be stored in a new ds-array that can be used to perform additional computations.

Dividing ds-arrays in blocks also provides performance improvements in other operations. One of these operations is the transpose, which can be computed by transposing each block in parallel, and then rearranging the list of lists of blocks in the ds-array. Transposing a ds-array thus requires one task per block, while transposing a Dataset requires $N^2 + N$ tasks, where $N$ is the number of Subsets.

Finally, using the collections data type to create tasks with a variable number of inputs and outputs improves the performance of operations that need to redistribute data, such as shuffling. More precisely, the number of tasks required for this type of operations is $2N$ with collections and $N^2 + N$ without collections, where $N$ is the number of partitions.

\section{Experimental evaluation}
\label{sec:evaluation}

In this section, we analyze the performance of ds-arrays compared to Datasets. In our analysis, we use two data operations and two machine learning models. The two data operations are the transpose and shuffle, and the models are K-means clustering and alternating least squares (ALS). We choose the transpose, shuffle and ALS because they represent different examples on how ds-arrays can improve performance over Datasets. We use K-means as an example of an algorithm that is not affected by the underlying data representation.

\subsection{Testbed}

Our experiments are run on the MareNostrum~4 supercomputer\footnote{https://www.bsc.es/marenostrum}. Each MareNostrum~4 node has two Intel Xeon Platinum 8160 24C at 2.1 GHz chips with 24 cores each, and 96GB of memory. Nodes run Linux, are interconnected through an Intel Omni-Path architecture, and have access to a shared GPFS storage system.

\subsection{Transpose}

The transpose is a common operation that is required to solve numerous problems. In dislib, the ALS algorithm transposes the input Dataset, which is viewed as a matrix, to be able to access both rows and columns in an efficient way. This is needed because Datasets do not provide efficient access to columns as discussed in Section~\ref{sec:limitations}. 

The transpose operation on a Dataset requires partitioning each Subset and merging the partitions to build new Subsets with transposed data. Given a Dataset with $N$ Subsets, this process requires $N^2$ tasks to divide each Subset into $N$ parts, and $N$ more tasks to merge these parts into new Subsets. In total, $N^2 + N$ tasks are required. The complexity of transposing a Dataset is caused by the need of maintaining data divided in Subsets.

In contrast to this, transposing a ds-array is less complex because it stores data in square blocks of an arbitrary size. Transposing a ds-array can be achieved by transposing each row of blocks, and then reorganizing the blocks in the ds-array such that block $(i, j)$ becomes block $(j, i)$. That is, given a ds-array with $N \times M$ blocks, the transpose requires only $N$ tasks, and the transposed ds-array has $M \times N$ blocks. 

\begin{figure*}[t!]
    \centering
    \begin{subfigure}[]{\textwidth}
	    \centering
        \includegraphics[width=0.4\textwidth]{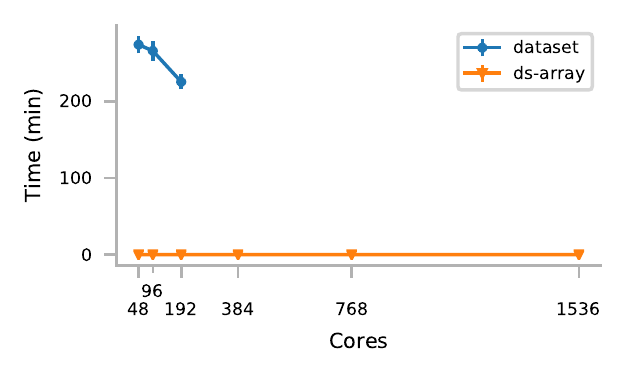}
        \includegraphics[width=0.4\textwidth]{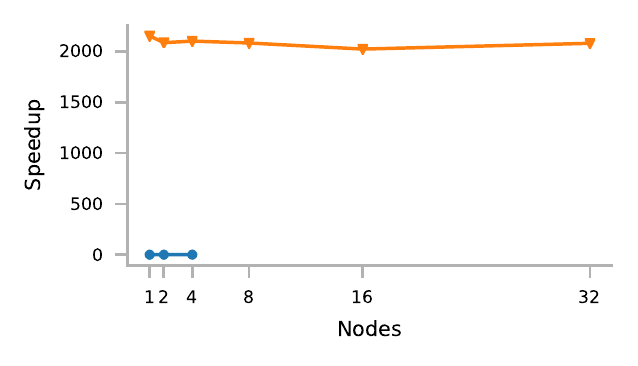}
        \caption{Strong scalability.}
    \end{subfigure}
    
    \begin{subfigure}[]{\textwidth}
	    \centering
        \includegraphics[width=0.4\textwidth]{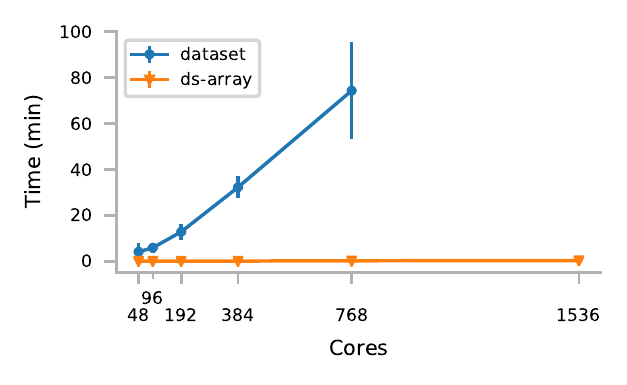}
        \includegraphics[width=0.4\textwidth]{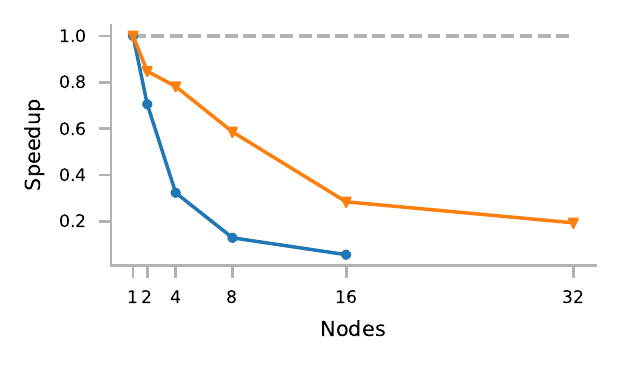}
        \caption{Weak scalability.}
    \end{subfigure}
    
	\caption{Execution times and speedup of transpose using Datasets and ds-arrays.}
	\label{fig:transpose}
\end{figure*}

The transpose operation is a good example on how the internal organization of Datasets can limit performance. Figure~\ref{fig:transpose} shows strong and weak scalability of the transpose operation using Datasets and ds-arrays. For strong scalability, we use 46,080 samples of 46,080 features, Datasets with 1,536 Subsets, and ds-arrays with 1,536$\times$1 blocks. We use the Dataset execution with 48 cores as baseline to compute speedup. For weak scalability, we use 500 samples per core with 100,000 features, Datasets with 1 Subset per core, and ds-arrays with $N_{cores} \times 1$ blocks (i.e., one block of rows per core). Reported times are the mean of 10 executions, and in the experiments with Datasets, missing data points are because of memory issues due to handling a large number of tasks and resources.

We see that there is a significant improvement in performance when using ds-arrays. More precisely, ds-arrays reduce execution times by up to 99\%. Going from 4.5 hours to 7 seconds in the strong scalability case, and from 1.5 hours to 14 seconds in the weak scalability case with 768 cores. In terms of speedup, we see very poor strong scalability both for Datasets and ds-arrays. In the case of ds-arrays, this is because execution times are too low to allow for any improvement when increasing the number of cores. Ds-arrays also show better weak scalability than Datasets. However, ds-arrays also show a performance degradation as we increase the number of cores. This is because PyCOMPSs scheduling overhead is proportional to the number of cores and tasks, and the transpose tasks have very fine granularity. Experiments with a bigger problem yield better scalability results for ds-arrays, but are intractable when using Datasets, and thus cannot be used in this performance comparison.

\subsection{Alternating least squares}

ALS is a matrix factorization method that has proven to be very useful for collaborative filtering~ref{zhuo08}, which can be applied to build recommendation systems. Dislib's current implementation of ALS treats samples in a Dataset as a matrix, and alternates accesses to its rows and columns. Since Datasets are divided in Subsets (i.e., by blocks of rows), access to arbitrary columns cannot be done efficiently. The current implementation of ALS creates a transposed copy of the input Dataset to be able to efficiently access its columns. This doubles the amount of memory used and introduces computation overhead. 

In contrast to this, ds-arrays provide efficient access to both rows and columns because ds-arrays can be partitioned in blocks of an arbitrary size. This eliminates the need for the transposed copy of the input ds-array. The ALS algorithm is another example on how a more convenient distributed data structure can improve performance and reduce computational complexity.

\begin{figure}[t!]
    \centering
    \includegraphics[width=0.4\textwidth]{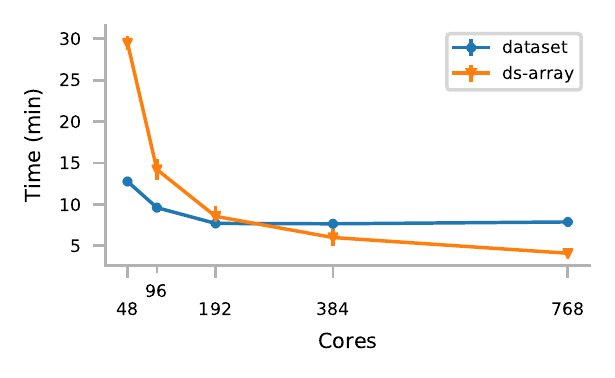}
    \includegraphics[width=0.4\textwidth]{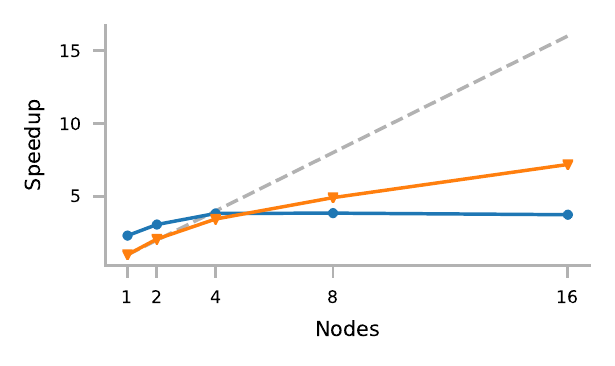}
	\caption{Execution times and speedup of ALS using Datasets and ds-arrays.}
	\label{fig:als}
\end{figure}

Figure~\ref{fig:als} shows execution times and speedup of ALS using Datasets and ds-arrays. We use the Netflix Prize data set as input~\cite{netflix}, which is a publicly available data set of movie ratings containing 17,770 rows (ratings) and 480,189 columns (users). The total number of ratings is 100,480,507  as the data set is extremely sparse. We use Datasets with 192 Subsets and ds-arrays of $192 \times 192$ blocks. Reported times correspond to the mean over 10 executions.

Ds-arrays show better scalability than Datasets, and are faster when using a large number of cores. This is because ALS does not need to perform a transpose when using ds-arrays. As seen before, the transpose operation has poor scalability in the number of resources. However, ds-arrays are slower than Datasets when using a small number of cores. The reason for this is the high number of blocks in ds-arrays compared to the number of Subsets in Datasets. We divide ds-arrays in $192 \times 192$ blocks, which results in 36,864 blocks in total, while Datasets are divided in just 192 Subsets. This enables efficient access to columns in the case of ds-arrays, but also introduces the overhead of handling a much larger number of partitions. This increases individual task scheduling time and can add up to minutes over the whole execution. 

\subsection{Shuffle}

The idea of the shuffle operation is to randomly redistribute samples across different Subsets. As this is extremely costly, the original implementation of shuffle in dislib performs a less expensive pseudo shuffle that is sufficient for most use cases. Given a Dataset with $N$ Subsets of size $S$, the shuffle algorithm splits each Subset into $N$ random parts, and creates $N$ new Subsets by merging one part from each original Subset. If the number of Subsets $N$ is greater than their size $S$, each original Subset is distributed among $S$ new Subsets in a way that the final shuffled $N$ Subsets are also of size $S$. The number of tasks required to shuffle a Dataset is $N \cdot min(N, S) + N$.

In contrast to this, shuffling the rows of a ds-array requires less tasks thanks to the use of PyCOMPSs collections (see Section~\ref{sec:internal}). The collection parameter is used to define multiple inputs or outputs in tasks, and thus, simplifies the process of redistributing one partition into $N$ new partitions. Given a ds-array of $N \times M$ blocks, pseudo shuffling its rows requires $2N$ tasks.

\begin{figure}[t!]
    \centering
    \includegraphics[width=0.4\textwidth]{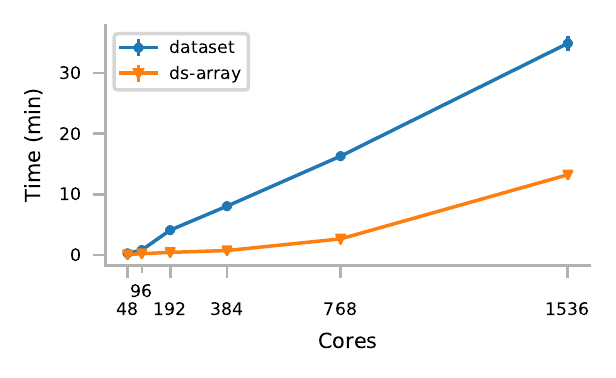}
    \includegraphics[width=0.4\textwidth]{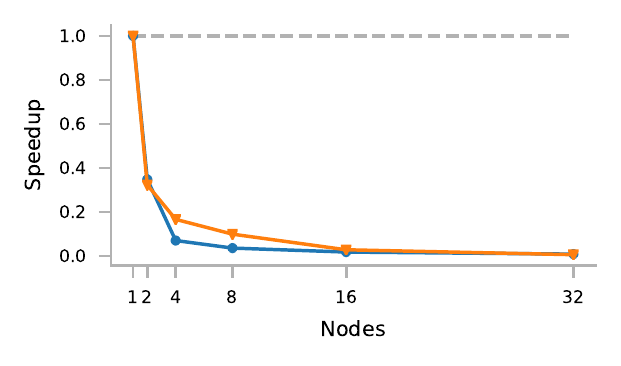}
	\caption{Weak scalability of shuffle using Datasets and ds-arrays.}
	\label{fig:shuffle}
\end{figure}

The best way to evaluate how the difference in the number of tasks affects performance is to analyze weak scalability as with the transpose operator. Figure~\ref{fig:shuffle} shows weak scalability of shuffle when using Datasets and ds-arrays. In the experiments, we use 300 samples of 2 features per core, Datasets with one Subset per core, and ds-arrays of $N_{cores} \times 1$ blocks. As in previous experiments, reported times are the mean of 10 executions.

We see that shuffle is much faster with ds-arrays than with Datasets as we increase the number of cores and partitions. This is consistent with the fact that the number of tasks when using Datasets grows much faster with the number of partitions than when using ds-arrays. When using 1,536 cores, the performance improvement is of around 60\%, and we expect this improvement to be more significant in scenarios with larger partitions and more number of cores.

Both Datasets and ds-array show limited speedup as the number of cores increases. This is because shuffle is a costly operation that grows in complexity as we increase the number of partitions, and because the operation consists of a large number of short tasks. This overloads PyCOMPSs' scheduler, especially when there is also a large number of computational resources.

\subsection{K-means}

\begin{figure}[h]
    \centering
    \includegraphics[width=0.4\textwidth]{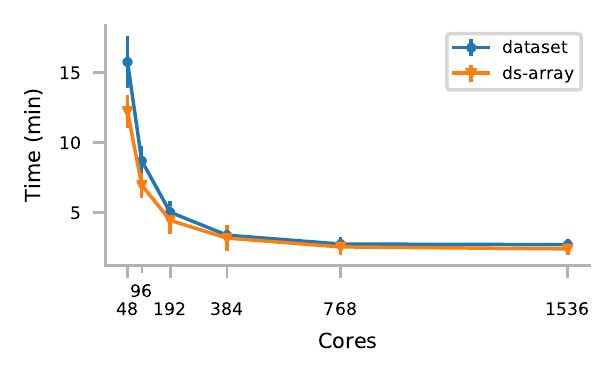}
    \includegraphics[width=0.4\textwidth]{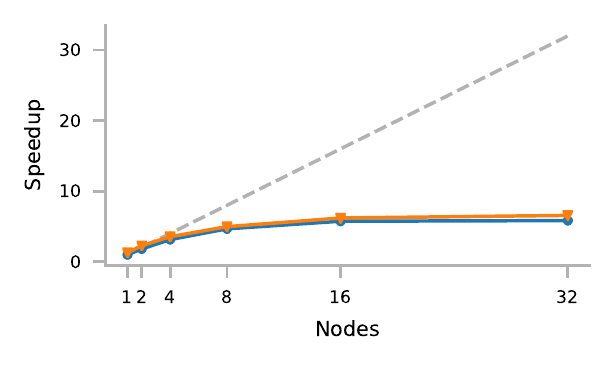}
	\caption{Strong scalability of K-means using Datasets and ds-arrays.}
	\label{fig:kmeans}
\end{figure}

K-means~\cite{hartigan79} is a well-known clustering algorithm. We use K-means as an example of an algorithm that is not affected by the underlying data representation and an algorithm that is used for large data sets~\cite{lu19}. Given a set of samples or data points, K-means first randomly places a predefined number of cluster centers, and then optimizes the location of these centers by computing the mean of the samples closer to each of them in an iterative process. The current implementation of K-means in dislib computes this mean in parallel by first computing partial sums on each Subset, and then carrying out a reduction. K-means with ds-arrays follows the same parallelization strategy, and thus, there should be no significant differences in performance between Datasets and ds-arrays when running K-means. Figure~\ref{fig:kmeans} shows execution times and speedup of K-means with Datasets and ds-arrays. We run the algorithm with around 50 million randomly generated samples of 1,000 features divided in 1,536 partitions.

We can see that the difference in performance between Datasets and ds-arrays is not significant. This is consistent with the fact that the improvements in ds-arrays do not have a direct impact on K-means due to the characteristics of the algorithm. Moreover, this shows that ds-arrays do not have a negative impact on performance as a result of changes in the internal organization of the data.

\subsection{Discussion}

Our experiments show that ds-arrays are more efficient than Datasets, and can improve execution time from hours to seconds in some cases. This performance improvement is due to a more flexible internal organization, as well as the use of collection parameters in tasks, which greatly simplify certain workflows. The differences in performance between ds-arrays and Datasets increase with the number of cores and data partitions. This means that ds-arrays scale better, and allow the processing of larger data sets with a larger number of resources. This is most noticeable in the transpose case, where the problem becomes unfeasible very fast using Datasets, but can be computed in seconds using ds-arrays.

Our experiments also show that performance decreases as we increase the number of data partitions both in Datasets and ds-arrays. This can outweigh ds-arrays' performance improvements that result from a more flexible data partitioning, like in the ALS case. However, this performance reduction is not as significant as the improvement obtained in other cases, and becomes irrelevant as we increase the number of cores. Moreover, as seen in the experiments with K-means, ds-arrays show similar performance to Datasets when using the same number of partitions in cases where the algorithm does not take advantage of ds-array's advantages. This means that ds-arrays achieve the same performance as Datasets in the worst case, while being two orders of magnitude faster in the best case.

\section{Conclusions}
\label{sec:conclusions}

In this paper, we present a new distributed data structure for dislib called distributed array (ds-array). Distributed arrays overcome several limitations of dislib's current main data structures (Datasets and Subsets), including the lack of usability and flexibility, a counter-intuitive API, and limited performance and scalability. Ds-arrays are a more generic data abstraction than Datasets and provide a NumPy-like API that is more intuitive to Python developers. Ds-arrays internal structure also makes them more flexible than Datasets, simplifies dislib's API, and improves performance in various ways. Our experiments show that ds-arrays can greatly speed up the computation of certain algorithms, and can reduce execution time from hours to seconds. This performance improvement is more significant as we increase the number of cores and data partitions, which makes ds-arrays a better suited data structure for dealing with large data sets in large scale distributed platforms.

Finally, ds-arrays extend dislib's functionality to common mathematical operations, such as matrix multiplication and decomposition, in a more natural way than using Datasets. This makes dislib a distributed version of the combination of NumPy and scikit-learn, and brings dislib closer to other existing approaches, like Dask and Dask-ML, with the advantage of providing better performance and scalability when dealing with extremely large data sets and a large number of computational resources~\cite{alvarez19}.

\section*{Acknowledgements}
This work has been supported by the Spanish Government (SEV2015-0493), by the Spanish Ministry of Science and Innovation (contract TIN2015-65316-P), by Generalitat de Catalunya (contract 2014-SGR-1051). The research leading to these results has also received funding from the collaboration between Fujitsu and BSC (Script Language Platform).

%%
%% The next two lines define the bibliography style to be used, and
%% the bibliography file.

\bibliographystyle{elsarticle-num}
\bibliography{main}

\end{document}